\documentclass{PoS}

\usepackage{amsfonts,amssymb,amsmath,bm}
\usepackage{graphicx}

\def\s#1{{\scriptscriptstyle #1}}

\def\diff{{\mathrm d}}
\def\mathi{{\mathrm i}}

\newcommand{\be}{\begin{equation}}
\newcommand{\bea}{\begin{eqnarray}}
\newcommand{\ee}{\end{equation}}
\newcommand{\eea}{\end{eqnarray}}

\title{IR properties of Yang-Mills theories from the Batalin-Vilkovisky formalism}

\ShortTitle{IR properties of QCD from the Batalin-Vilkoviski formalism}

\author{\speaker{Daniele Binosi}\\
        European Centre for Theoretical Studies in Nuclear Physics and Related Areas (ECT*) and Fondazione Bruno Kessler,\\ Villa Tambosi, Strada delle Tabarelle 286, I-38123 Villazzano (TN), Italy\\
        E-mail: \email{binosi@ect.it}}

\abstract{The powerful quantization formalism of Batalin and Vilkovisky streamlines the derivation of the complete set of (non-linear) identities arising from the local BRST symmetry of Yang-Mills theories.  When applied in the Background Field Method type of gauges, it also gives rise to identities which relate Green's functions involving background fields to Green's functions involving quantum fields. All these identities lie at the core of the recent progress in understanding from the continuum formulation  the IR dynamics emerging from lattice simulations. In this talk, we will first review the Batalin-Vilkovisky formalism and then apply it to the problem of extracting the  effective charge from the available lattice data.}

\FullConference{The many faces of QCD\\
		November 2-5, 2010\\
		Gent Belgium}

\begin{document}

\section{Introduction}

\noindent The Batalin-Vilkovisky (BV) formalism~\cite{Batalin:1977pb} is a very powerful quantization framework introduced long ago to deal with the quantization of very general gauge theories, including those with reducible or open symmetry algebras ({\it e.g.}, certain formulations of supergravity). There are many areas in which this formalism has proved invaluable, and the Pinch Technique (PT)~\cite{Cornwall:1982zr} is one of those.
Indeed the application of the BV method in this context~\cite{Binosi:2002ez} has allowed the PT to transcend its diagrammatic origins, to become a fully fledged formal tool capable of enforcing explicit gauge invariance in (all-order) off-shell Green's functions~\cite{Binosi:2002ft} and the Schwinger-Dyson equations that couples them~\cite{Binosi:2007pi}, as well as reproducing in an elegant and compact way~\cite{Aguilar:2008xm} the recent large volume lattice data revealing an IR finite gluon propagator and ghost dressing function~\cite{Cucchieri:2007md}.

In this talk, we will review the BV formalism within the $SU(N)$ Yang-Mills theories, placing particular emphasis on how it streamlines the derivation of the complete set of identities -- Slavnov-Taylor identities (STIs) in the case of the conventional $R_\xi$ gauges and, in addition, background-quantum identities (BQIs) and Ward identities (WIs) when applied to Background Field Method (BFM) type of gauges --  arising from the local BRST symmetry. Through the derivation and calculation of the Yang-Mills effective charge we will also show that, when properly combined, these identities provide invaluable information about the underlying IR dynamics.

\section{Batalin-Vilkovisky formalism: a primer}

\noindent As everybody knows, the classical action of a $SU(N)$ Yang-Mills theory\footnote{We concentrate for convenience on the case of pure Yang-Mills theories; the inclusion of fermions does not present any problem.} is invariant under the BRST transformations 
\be
s A^m_\mu=({\cal D}_\mu c)^m; \qquad
s c^m=-\frac12gf^{mnr}c^nc^r;\qquad
s\bar c^m=B^m;\qquad sB^m=0,
\ee
where $s$ is the BRST operator, $({\cal D}_\mu c)^m=\partial_\mu c^m+gf^{mnr}A^n_\mu c^r$ is the usual covariant derivative, while $B$ represents the so-called Nakanishi-Lautrup multiplier corresponding to a yet to be specified gauge-fixing condition~${\cal F}$. 

An efficient method for elevating this symmetry to the quantum level
is by applying the aforementioned BV method~\cite{Batalin:1977pb}, which starts by introducing an anti-field $\Phi^*$ for each field $\Phi$ which transforms non-linearly under the BRST operator. The anti-fields $\Phi^*$ have opposite  statistics with respect   to  the corresponding fields $\Phi$, a ghost charge  ${\rm gh}(\Phi^*)= {\rm gh}(\Phi^*)=-1- {\rm gh}(\Phi)$ and, choosing the (mass) dimension of the ghost field to be 0, a dimension $\mathrm{dim}(\Phi^*)=4-\mathrm{dim}(\Phi)$.

The next step is  to add to the original (gauge fixed) action the coupling term $\sum\Phi^*s\ \Phi$,  so that it now reads ($\xi$ is the gauge fixing parameter)
\bea
\Gamma^{(0)}&=&\int\!\mathrm{d}^4x\left[-\frac14F^m_{\mu\nu} F^{\mu\nu}_m+{\cal L}_{\s{\mathrm{GF}}}+{\cal L}_{\s{\mathrm{FPG}}}+A^{*m}_\mu\left({\cal D}^\mu c\right)^m-\frac12gf^{mnr} c^{*m}c^nc^r\right] \nonumber \\
{\cal L}_{\s{\mathrm{GF}}}+{\cal L}_{\s{\mathrm{FPG}}}&=&s\left(\bar c^m\mathcal{F}^m-\frac\xi2\bar c^m B^m\right).
\label{clact}
\eea
Then, the original gauge invariance of the theory plus the nihilpotency of the BRST operator, makes it relatively easy to prove that the classical action above satisfies the master equation
\begin{equation}
\int\!\diff^4x\left\{\frac{\delta\Gamma^{(0)}}{\delta A^{*\mu}_m}\frac{\delta\Gamma^{(0)}}{\delta A^{m}_\mu}+\frac{\delta\Gamma^{(0)}}{\delta c^{*m}}\frac{\delta\Gamma^{(0)}}{\delta c^{m}}+B^m\frac{\delta\Gamma^{(0)}}{\delta\bar c^m}\right\} = 0.
\label{master_eq}
\end{equation}
Now, the BRST symmetry is crucial for endowing a theory  with a unitary $S$-matrix and gauge-independent physical observables; therefore, one implements it to the all-order level by establishing the quantum corrected version of the master equation (\ref{master_eq}) in the form of the complete STI functional
\be
{\cal S}_{\s{\mathrm C}}(\Gamma)[\Phi]=\int\!\diff^4x\left\{\frac{\delta\Gamma}{\delta A^{*\mu}_m}\frac{\delta\Gamma}{\delta A^{m}_\mu}+\frac{\delta\Gamma}{\delta c^{*m}}\frac{\delta\Gamma}{\delta c^{m}}+B^m\frac{\delta\Gamma}{\delta\bar c^m}\right\} = 0,
\label{STIfunc_nm}
\ee
where $\Gamma=\Gamma[\Phi,\Phi^*]$ is now the effective action. 

When dealing with {\it linear} gauge fixing functions ${\cal F}$, such as the conventional $R_\xi$ gauge \mbox{${\cal F}^m=\partial^\mu A^m_\mu$}, the structure of the STI generating functional of Eq.~(\ref{STIfunc_nm}) can be further simplified  by omitting the last term proportional to the $B$ field; we thus obtain the {\it reduced} STI functional  
\be
{\cal S}(\Gamma)[\Phi]=\int\!\diff^4x\left\{\frac{\delta\Gamma}{\delta A^{*\mu}_m}\frac{\delta\Gamma}{\delta A^{m}_\mu}+\frac{\delta\Gamma}{\delta c^{*m}}\frac{\delta\Gamma}{\delta c^{m}}\right\} = 0.
\label{STIfunc}
\ee
In practice, the STIs generated from this {\it reduced} functional  coincide with the ones that would have been obtained from the complete functional after the implementation of the so-called ghost (or Faddeev-Popov) equation described  in Section~\ref{FPEs} below~\cite{Itzykson:1980rh}. 

The STI functional of Eq.~(\ref{STIfunc_nm}) can be easily adapted to the BFM type of gauges, where one splits the gluon field into a background $\widehat{A}$ and a quantum part, performs the shift $A\to A+\widehat{A}$ and retains gauge invariance with respect to the background field by choosing the special gauge fixing 
\be
{\cal F}^m=(\widehat{\cal D}^\mu A_\mu)^m=\partial^\mu A^m_\mu+gf^{mnr}\widehat{A}_n^\mu A^r_\mu,
\ee
($\widehat{\cal D}$ is the background covariant derivative). In order to implement the equations of motion for the background fields at the quantum level, one next  extends the BRST symmetry to the background gluon field, through the relations
\begin{equation}
s\widehat{A}_\mu^m=\Omega^m_\mu, \qquad s\Omega^m_\mu=0,
\label{extBRST}
\end{equation}
with $\Omega$ denoting a (classical) vector field with the same quantum numbers as the gluon, ghost charge $+1$ and Fermi statistics. The dependence of the Green's  functions on the background fields is then controlled by  the modified STI functional
\be
{\cal S}'(\Gamma')[\Phi]={\cal S}(\Gamma')[\Phi]+\int\!\diff^4x\
\Omega_m^\mu\left(\frac{\delta\Gamma'}{\delta\widehat{A}^m_\mu}-\frac{\delta\Gamma'}{\delta A^m_\mu}\right)=0,
\label{STIfunc_BFM}
\ee
where $\Gamma'$ denotes the effective action that depends on the background source $\Omega$ (with $\Gamma\equiv\Gamma'\vert_{\Omega=0}$), and ${\cal S}(\Gamma')[\Phi]$ is the reduced STI functional of Eq.~(\ref{STIfunc}) (since the BFM gauge fixing function is linear in the quantum field, we can indeed restrict our considerations to the reduced STI functional alone also in this case).  The functional (\ref{STIfunc_BFM}) above will then provide the BQIs, which as already mentioned, relate 1PI Green's functions involving background fields with the ones involving quantum fields. 

Finally, the background gauge invariance of the BFM effective action is encoded into the WI functional
\be
{\cal W}_{\vartheta}[\Gamma']=\int\!\diff^4x\,\sum_{\varphi=\Phi,\Phi^*}\left(\delta_{\vartheta}\varphi\right)\frac{\delta\Gamma'}{\delta\varphi}=0,
\label{WI_gen_funct}
\ee 
where $\vartheta^m$ (that now plays the role of the ghost field) is the local infinitesimal parameter corresponding to the $SU(N)$ generators $t^m$. The transformations $\delta_{\vartheta}\Phi$ are thus given by
\bea
\delta_{\vartheta}A^m_\mu=gf^{mnr}A^n_\mu \vartheta^r &\qquad& \delta_{\vartheta}\widehat{A}^m_\mu=\partial_\mu \vartheta^m+gf^{mnr}\widehat{A}^n_\mu \vartheta^r,\nonumber \\
\delta_{\vartheta} c^m=-g f^{mnr}c^n\vartheta^r &\qquad& \delta_{\vartheta} \bar c^m=-g f^{mnr}\bar c^n\vartheta^r,
\label{theta_trans}
\eea
and the corresponding anti-fields transformations $\delta_{\vartheta}\Phi^*$ coinciding with the  transformations of the corresponding quantum fields above according to their specific representations. This functional will give rise to the WIs satisfied by Green's functions when contracted with the momentum corresponding to a background leg.

\subsection{Functional differentiation rules}

\noindent Slavnov-Taylor, Background-Quantum and Ward identities are all obtained by taking functional derivatives of the corresponding functionals ${\cal S}$, ${\cal S}'$ and ${\cal W}$ [Eqs.~(\ref{STIfunc}), (\ref{STIfunc_BFM}) and (\ref{WI_gen_funct}) respectively], setting afterwards  all fields, anti-fields and sources to zero.
However, in order to reach meaningful expressions, one needs to keep in mind that:
\begin{enumerate} 
\item ${\cal S}$ and ${\cal S}'$ have ghost charge 1; 
\item Functions with non-zero ghost charge vanish, for the ghost charge is a conserved quantity.
\end{enumerate}
Then in order to extract non-zero identities the following rules apply
\begin{itemize}

\item {\it Slavnov-Taylor identities}. In this case one needs to differentiate the functional ${\cal S}$ of Eq.~(\ref{STIfunc}) with respect to a  combination of fields, containing either one ghost field, or two ghost fields and one anti-field. The only exception to this rule is when differentiating with respect to a ghost anti-field, which needs to be compensated by three ghost fields. In particular, identities involving one or more gauge fields are obtained by differentiating ${\cal S}$
with respect to the set of fields in which one gauge boson has been replaced by the corresponding ghost field. This is due to the fact that  the linear part of the BRST transformation of the gauge field is proportional to the ghost field: $s\,A^m_\mu|_\mathrm{linear}=\partial_\mu c^m$. For completeness we notice that, for obtaining STIs involving Green's functions that contain ghost fields, one ghost field must be replaced by two ghost fields, due to the non linearity of the corresponding BRST transformation ($s\,c^m\propto f^{mnr}c^nc^r$). This implies in turn that only certain (properly symmetrized) combinations of ghost Green's functions will appear in these identities, sometimes limiting their usefulness.

\item {\it Background-Quantum identities}. In this case the rule is very simple, since all one needs to do is to differentiate the STI functional ${\cal S}'$ of Eq.~(\ref{STIfunc_BFM}) with respect to the background source $\Omega$ and the needed combination of fields (background or quantum).

\item{\it Ward identities}.  Finally, in order to obtain the WIs satisfied by the Green's functions involving background gluons $\widehat{A}$, one has to differentiate the functional ${\cal W}$ of Eq.~(\ref{WI_gen_funct}) with respect to a combination of fields in which the background gluon has been replaced by the corresponding gauge parameter $ \vartheta$. 

\end{itemize}

The last technical point to be clarified is the dependence of the identities on the (external) momenta. After Fourier transforming the result of the differentiation, one should notice that the integral over $\diff^4x$ appearing in Eqs.~(\ref{STIfunc}), ~(\ref{STIfunc_BFM}) and~(\ref{WI_gen_funct}), together with the conservation of momentum flow of the Green's functions, implies that no momentum integration is left over; as a result, the identities will be expressed as a sum of products of (at most two) Green's functions.

\subsection{Ghost and anti-ghost equations\label{FPEs}}

\noindent There are two more useful equations that can be written down. To do that, let us start notice that the (BFM) equation of motion of the $B$ field reads
\be
\frac{\delta\Gamma}{\delta B^m}-(\widehat{\cal D}^\mu A_\mu)^m=0,
\ee
(for the equivalent equation in the $R_\xi$  gauges just set the background field $\widehat{A}$ to zero). This equation in conjunction with the linearity of the gauge fixing function, implies the existence of a constraint that takes the form of the so-called ghost (or Faddeev-Popov) equation
\begin{equation}
\frac{\delta\Gamma'}{\delta \bar c^m}+\left(\widehat{{\cal D}}^\mu\frac{\delta\Gamma'}{\delta A^*_\mu}\right)^m
-\left({\cal D}^\mu\Omega_\mu\right)^m-gf^{mrs}\widehat{A}^r_\mu\Omega^\mu_s=0,
\label{FPeqBFM}
\end{equation}
(again for the equivalent equation in the $R_\xi$ gauges set the background field $\widehat{A}$ and source $\Omega$ to zero). Notice that by ``undoing'' the splitting of the field $A$ into background and quantum parts (that is using $A+\widehat{A}\to A$) we can write the equation above in the more compact form
\be
\frac{\delta\Gamma'}{\delta \bar c^m}+\left(\widehat{{\cal D}}^\mu\frac{\delta\Gamma'}{\delta A^*_\mu}\right)^m
-\left({\cal D}^\mu\Omega_\mu\right)^m=0.
\label{FPeqBFM-1}
\ee

Finally, when considering the background-Landau gauge $(\widehat{\cal D}^\mu A_\mu)^m=0$, the additional (local) anti-ghost equation appears, that reads~\cite{Grassi:2004yq}
\be
\frac{\delta\Gamma'}{\delta c^m}-\left({\cal D}^\mu A^*_\mu\right)^m
-\left(\widehat{{\cal D}}^\mu\frac{\delta\Gamma'}{\delta\Omega_\mu}\right)^m+gf^{mnr}\frac{\delta\Gamma}{\delta B^n}\bar c^r-gf^{mnr}c^{*n}c^r=0.
\label{AGE}
\ee
Notice that in the conventional $R_\xi$ Landau gauge only an integrated (and correspondingly less powerful) version of this identity exists.

\section{Examples}

\noindent In this section we give several examples of the kind of interesting and powerful identities that can be obtained within the BV framework introduced before. The important point that we want to stress is that every auxiliary function that appears in the identities below can be explicitly calculated by using the set of Feynman rules derived from the action (\ref{clact}) -- see for example the second paper in~\cite{Binosi:2007pi}.

\subsection{Background-quantum identities in the gluon two-point sector}

\noindent Consider first the two-point gluon sector. Differentiating the STI functional ${\cal S}'$ of Eq.~(\ref{STIfunc_BFM}) with respect to the combinations consisting of a background source/field and a background source/quantum field. On then gets the following identities
\bea
\mathi\Gamma_{\widehat{A}^m_\mu A^n_\nu}(q)&=&\left[\mathi g_\mu^\rho\delta^{mr}+\Gamma_{\Omega^m_\mu A^{*\rho}_r}(q)\right]\Gamma_{A^r_\rho A^n_\nu}(q)\nonumber \\
\mathi\Gamma_{\widehat{A}^m_\mu \widehat{A}^n_\nu}(q)&=&\left[\mathi g_\mu^\rho\delta^{mr}+\Gamma_{\Omega^m_\mu A^{*\rho}_r}(q)\right]\Gamma_{A^r_\rho \widehat{A}^n_\nu}(q).
\eea
These two equations can be now combined in such a way that the two-point function mixing background and quantum fields drops out, to get the BQI\footnote{We are using here implicitly the transversality of the gluon two-point function.}
\be
\mathi\Gamma_{\widehat{A}^m_\mu \widehat{A}^n_\nu}(q)=\mathi\Gamma_{A^m_\mu A^n_\nu}(q)+2\Gamma_{\Omega^m_\mu A^{*\rho}_r}(q)\Gamma_{A^r_\rho A^n_\nu}(q)-\mathi\Gamma_{\Omega^m_\mu A^{*\rho}_r}(q)\Gamma_{A^r_\rho A^s_\sigma}(q)\Gamma_{\Omega^n_\nu A^{*\sigma}_s}(q).
\ee
At this point we introduce the gluon propagator as
\be
\mathi\Delta^{mn}_{\mu\nu}(q)=-\mathi\delta^{mn}\left[ P_{\mu\nu}(q)\Delta(q^2) +\xi\frac{q_\mu q_\nu}{q^4}\right]; \qquad P_{\mu\nu}(q)=g_{\mu\nu}-\frac{q_\mu q_\nu}{q^2}
\ee
and similarly for the background propagator $\widehat{\Delta}$; then,
making use of the decomposition
\be
\Gamma_{\Omega^m_\mu A^{*n}_\nu}(q)=\mathi\delta^{mn}\left[G(q^2)g_{\mu\nu}+\frac{q_\mu q_\nu}{q^2}L(q^2)\right], 
\label{GandL}
\ee
and the relations
\be
\Gamma_{A^m_\mu A^n_\nu}(q)=\mathi\delta^{mn}P_{\mu\nu}(q)\Delta^{-1}(q^2);\qquad
\Gamma_{\widehat{A}^m_\mu \widehat{A}^n_\nu}(q)=\mathi\delta^{mn}P_{\mu\nu}(q)\widehat{\Delta}^{-1}(q^2)
\ee
we arrive at the well known PT-BFM identity
\be
\Delta(q^2)=\left[1+G(q^2)\right]^2\widehat{\Delta}(q^2).
\label{BQI}
\ee
The above identity which represents the basic equation from which a gauge invariance truncation scheme for the Schwinger-Dyson equation of the gluon propagator can be derived~\cite{Binosi:2007pi}.

\subsection{Two-point ghost sector in the background Landau gauge\label{tpg}}

\noindent Next, let us now consider the ghost two-point sector in the background Landau gauge. Differentiating the ghost equation (\ref{FPeqBFM-1}) with respect to a ghost field and a background source we get the two relations
\bea
\Gamma_{c^m\bar c^n}(q)&=&-\mathi q^\nu\Gamma_{c^m A^{*n}_\nu}(q)\nonumber \\
\Gamma_{\bar c^n\Omega^m_\mu}(q)&=&q_\mu\delta^{mn}-\mathi q^\nu\Gamma_{\Omega_\mu^m A^{*n}_\nu}(q).
\label{FPE_ghost}
\eea
On the other hand, differentiation of the anti-ghost equation~(\ref{AGE}) with respect to a gluon anti-field and an anti-ghost, gives
\bea
\Gamma_{c^m A^{*n}_\nu}(q)&=&q_\nu\delta^{mn}-\mathi q^\mu\Gamma_{\Omega^m_\mu A^{*n}_\nu}(q)
\nonumber \\
\Gamma_{c^m\bar c^n}(q)&=&-\mathi q^\mu\Gamma_{\bar c^a\Omega^m_\mu}(q).
\label{AGE-1}
\eea
Contracting the first equation in~(\ref{AGE-1}) with $q^\nu$, and making use of  the first equation in~(\ref{FPE_ghost}), 
we see that the dynamics of the ghost sector is entirely encoded in the $\Gamma_{\Omega A^*}$ auxiliary function, since 
\be
\Gamma_{c^m\bar c^n}(q)=-\mathi q^2\delta^{mn}-q^\mu q^\nu\Gamma_{\Omega^m_\mu A^{*n}_\nu}(q).
\label{funrelbv}
\ee
Introducing  finally the ghost dressing function $F$ and the Lorentz decompositions
\be
\mathi D^{mn}(q^2)=\mathi \delta^{mn}\frac{F(q^2)}{q^2}; \qquad \Gamma_{c^mA^{*n}_\nu}(q)=q_\nu\delta^{mn} C(q^2);\qquad \Gamma_{\bar c^m \Omega^n_\nu}(q)=q_\nu\delta^{mn} E(q^2)
\ee
with $D$  the ghost propagator, we find that when combining Eqs.~(\ref{funrelbv}) and~(\ref{GandL})  with the last equation of~(\ref{FPE_ghost}) and~(\ref{AGE-1}) one gets  the identities~\cite{Grassi:2004yq}
\be
C(q^2) = E(q^2)\ =\ F^{-1}(q^2); \qquad
F^{-1}(q^2)=1+G(q^2)+L(q^2).
\label{ids}
\ee 

Now under very general conditions it can be proved that $L(0)=0$ so that one would get an IR divergent ghost dressing function if \mbox{$G(0)=-1$}. The latter condition reminds of the so-called Kugo-Ojima confinement criterion~\cite{Kugo:1979gm}, and indeed one has the equality~\cite{Grassi:2004yq,Aguilar:2009pp}
\be
u(q^2)=G(q^2);\qquad P_{\mu\nu}(q)\delta^{mn}u(q^2)=\frac{q_\mu q_\nu}{q^2}+\int\!\diff^4x\ \mathrm{e}^{-\mathi q\cdot(x-y)}\left\langle T\big[\left({\cal D}_\mu c\right)_x^m\left({\cal D}_\mu \bar c\right)_y^n\big]\right\rangle.
\label{GeqL}
\ee

The identity above is striking for not only it bridges two widely different approaches, namely the PT-BFM and the local covariant operator formalism, but it does it by linking the two functions -- $G$ and $u$ -- that play a central role in their respective frameworks. We conclude by observing that since the $G$ form factor can be determined to a good approximation from the available lattice data for the gluon and ghost propagators, Eq.~(\ref{GeqL}) allows for a comparison with direct lattice calculations of $u$~\cite{Sternbeck:2006rd}, and thus for an overall consistency check of the lattice results~\cite{Aguilar:2009pp}.

\subsection{Three-point auxiliary ghost sector}

\noindent Consider finally the differentiation of the STI functional~(\ref{STIfunc}) with respect to the field combination~$ccA^*$, and of the WI functional~(\ref{WI_gen_funct}) with respect to the combination $\vartheta cA^*$; one gets then the identities
\bea
\Gamma_{c^m A^{*\mu}_s}(q)\Gamma_{c^r A^s_\mu A^{*n}_\nu}(r,q,p)&=&
\Gamma_{c^rA^{*\mu}_s}(r)\Gamma_{c^m A^s_\mu A^{*n}_\nu}(q,r,p)-\Gamma_{c^mc^rc^{*s}}(q,r,p)\Gamma_{c^s A^{*n}_\nu}(p) \nonumber \\
q^\mu\Gamma_{c^r\widehat{A}^m_\mu A^{*n}_\nu}(r,q,p)&=&gf^{rms}\Gamma_{c^s A^{*n}_\nu}(p)-gf^{snm}\Gamma_{c^r A^{*s}_\nu}(r),
\eea
where $p+q+r=0$. 
These identities can be further simplified by making use of the ghost equation (\ref{FPeqBFM-1}) to write~\cite{Binosi:2007pi}
\be
\Gamma_{c^a A^{*b}_\nu}(q)=\delta^{ab}q_\nu F^{-1}(q^2),
\ee
and therefore 
\bea
q^\mu\Gamma_{c^r A^m_\mu A^{*n}_\nu}(r,q,p)&=&F^{-1}(r^2)r^\mu\Gamma_{c^m A^r_\mu A^{*n}_\nu}(q,r,p)-F^{-1}(p^2)p_\nu\Gamma_{c^mc^rc^{*n}}(q,r,p)\nonumber \\
q^\mu\Gamma_{c^r\widehat{A}^m_\mu A^{*n}_\nu}(r,q,p)&=&gf^{rmn}p_\nu F^{-1}(p^2)-gf^{rnm}r_\nu F^{-1}(r^2).
\label{exoc}
\eea

Though the functions $\Gamma_{cAA^*}$ and $\Gamma_{c\widehat{A}A^*}$  might appear exotic at a first sight, in fact they are not: they correspond to the auxiliary ghost Green's functions that are bound to appear in the STIs and the WI satisfied by the background-quantum-quantum gluon vertex $\Gamma_{\widehat{A}AA}$. The latter is the vertex that appears in the Schwinger-Dyson equation of the PT-BFM gluon propagator, and that plays a pivotal role for achieving the dynamical generation of a gluon mass~\cite{Aguilar:2008xm}.  In particular, the identities~(\ref{exoc}) are instrumental when trying to write this vertex in the most general form consistent with the STIs and WI it must satisfy\footnote{In fact, the first identity in Eq.~(\ref{exoc}) appears already in the classic paper~\cite{Ball:1980ax}, though the authors just proved it at the one-loop level.}~\cite{Joannis}. 

\section{Application: the QCD effective charge}

\noindent Until this point the discussion has been on a rather formal level and one might wonder  if the BV formalism (together with the identities it gives rise to) bears any phenomenologically relevance especially as far as the IR sector of Yang-Mills theories is concerned. The answer to this question is indeed in the affirmative, and to substantiate this claim we will devote the last part of this talk to define the notion of a renormalization group (RG) invariant effective charge~\cite{Binosi:2002vk,Aguilar:2008fh} and determine it numerically from the available lattice data~\cite{Aguilar:2010gm}.

\subsection{Definition}

\noindent The RG invariant Yang Mills effective charge represents a quantity that lies at the interface between perturbative and non-perturbative effects in QCD, providing a continuous interpolation between two physically distinct regimes: the deep UV, where perturbation theory is reliable, and the deep IR, where non-perturbative techniques must be employed. 

There are two possible RG invariant products  which can be used as a basis for the definition of the effective charge:
\begin{itemize}
\item $\widehat{r}(q^2)$ which exploits the non-renormalization property of the ghost vertex in the Landau gauge (and therefore is the common choice adopted by lattice practitioners) {\it i.e.,} the renormalization constants identity $Z_gZ_A^{1/2}Z_c=1$; 
\item $\widehat{d}(q^2)$ which exploits the fact that PT-BFM quantities  satisfy WIs (as opposed to the usual STIs)  which lead to the renormalization constants identity $Z_gZ_{\widehat{A}}^{1/2}=1$.
\end{itemize}
Specifically, one has
\be
\widehat{r}(q^2)=g^2(\mu^2)\Delta(q^2)F^2(q^2); \qquad \widehat{d}(q^2)=g^2(\mu^2)\widehat{\Delta}(q^2)=g^2(\mu^2)\frac{\Delta(q^2)}{[1+G(q^2)]^2}
\label{RGIs}
\ee
where in the last equation the BQI (\ref{BQI}) was employed.
These two {\it dimensionful} quantities, that have a mass dimension of $-2$, share an important common ingredient, namely the scalar cofactor of the gluon propagator $\Delta$ which actually sets the scale. The next step is then to extract a {\it dimensionless} quantity that would correspond to the nonperturbative effective charge. Perturbatively, {\it i.e.}, for asymptotically large momenta, it is clear that the mass scale is saturated simply by $q^2$, the bare gluon propagator, and the effective charge is defined by pulling a $q^{-2}$ out of the corresponding RG invariant quantity. However,  in the case of a dynamically generated gluon mass, the gluon propagator becomes effectively massive; therefore, particular care is needed in deciding exactly what combination of mass scales ought to be pulled out. The correct procedure in such a case is to pull out a massive propagator of the form (in Euclidean space) $[q^2+m^2(q^2)]^{-1}$, with $m^2(q^2)$ the dynamical gluon mass\footnote{Given that the gluon propagator is finite in the IR, if one insists on factoring out a simple $q^{-2}$ term, one would get a completely unphysical coupling, namely, one that vanishes in the deep IR, where QCD is expected to be (and is) strongly coupled.}~\cite{Cornwall:1982zr}.
We then have~\cite{Aguilar:2008fh}
\be
\alpha_{\mathrm{gh}}(q^2)=\alpha(\mu^2)[q^2+m^2(q^2)]\Delta(q^2)F^2(q^2);\qquad 
\alpha(q^2)=\alpha(\mu^2)[q^2+m^2(q^2)]\frac{\Delta(q^2)}{[1+G(q^2)]^2}.
\label{echs}
\ee

In addition, due to the last identity of Eq.~(\ref{ids}), we can relate the two effective charges through~\cite{Aguilar:2008fh}
\be
\alpha(q^2)=\alpha_{\mathrm{gh}}(q^2)\left[1+\frac{L(q^2)}{1+G(q^2)}\right]^2.
\label{alphaeq}
\ee
Since $L(0)=0$ -- and $G(0)\neq-1$~\cite{Aguilar:2009pp} -- we therefore see that not only the two effective charges coincide in the UV region where they  reproduce the perturbative result, but also in the deep IR where one has \mbox{$\alpha(0)=\alpha_{\mathrm{gh}}(0)$}. 

\subsection{Numerics}

\begin{figure}[!t]
\includegraphics[scale=.95]{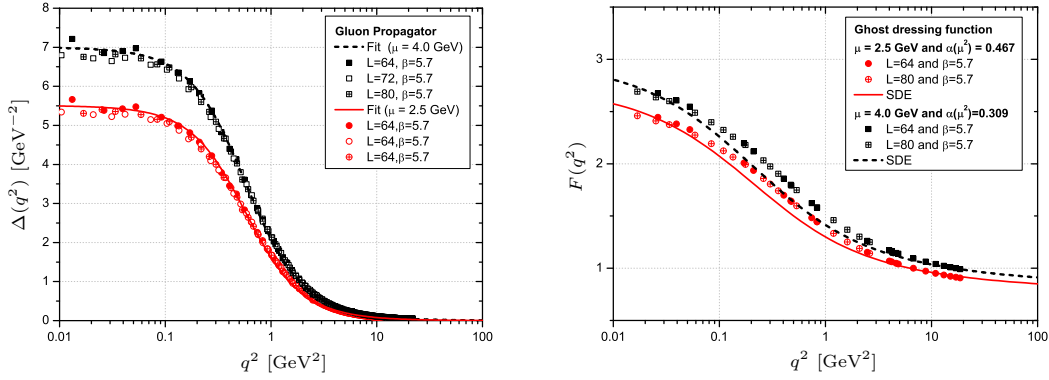}
\caption{\label{fig1}
{\it Left panel}: Lattice results for the gluon propagator scalar cofactor $\Delta$ at two different renormalization points $\mu$. {\it Right panel}: 
Comparison between the ghost dressing function $F$ obtained from the ghost Scwinger-Dyson equation  (lines) and the corresponding lattice data (points); as a result of this procedure one fixes $\alpha(\mu^2)= 0.467$ for $\mu = 2.5$ GeV and $\alpha(\mu^2)= 0.309$ for $\mu = 4.0$ GeV.}
\end{figure}

\noindent From the equations above it is clear that in order to calculate the effective charge we need knowledge of $\Delta$, $D$ (or equivalently $F$), $G$, the coupling $g$ and a suitable model for the running of the gluon mass (for details about the latter we refer to~\cite{Aguilar:2010gm}); on the other hand, $G$ (and $L$) can be expressed entirely in terms of $\Delta$, $D$ and $g$ by approximating the vertices appearing in their general expressions, with their tree-level values. One has then
\bea
G(q^2) &=& \frac{g^2 N}{3}\int\!\frac{\diff^4k}{(2\pi)^4} \left[2+ \frac{(k \cdot q)^2}{k^2 q^2}\right]\Delta (k)  D(k+q)
\nonumber\\
L(q^2) &=& \frac{g^2 N}{3}\int\!\frac{\diff^4k}{(2\pi)^4} \left[1 - 4 \, \frac{(k \cdot q)^2}{k^2 q^2}\right]\Delta (k)  D(k+q),
\label{simple}
\eea 
with $N$ the number of colors. 

The lattice gluon propagator is then taken as an input in our calculations; to determine the coupling, we instead solve the ghost dressing function Schwinger-Dyson equation for different values of $g$, fixing it to the value at which the best possible agreement with lattice results is reached\footnote{Obviously one must check that the coupling so obtained (at the renormalization scale used for the computation) is fully consistent with known perturbative results. In this case we checked against the 4-loop results of~\cite{Boucaud:2005rm} finding very good agreement~\cite{Aguilar:2010gm}.} (Fig.~\ref{fig1}).  

\begin{figure}[!t]
\includegraphics[scale=.95]{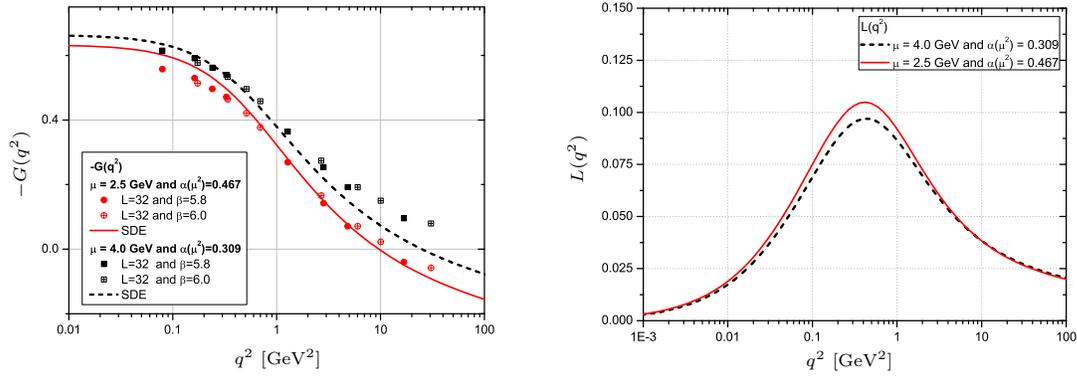}
\caption{\label{fig2}
{\it Left panel}: Comparison between the $G$ form factor (changed in sign) obtained from Eq.~(4.4) (continuous lines) and the corresponding lattice data (points) at two different renormalization points. Notice that the comparison is at most suggestive due to differences in the renormalization procedure; however one can clearly see that $-G$ saturates in the deep IR to a value much lower than 1 (around 0.6) and so the confinement criterion of Kugo-Ojima is not satisfied~\cite{Aguilar:2009pp}. {\it Right panel}: The $L$ form factor determined from Eq.~(4.4) at the same renormalization points. Observe the relative suppression of $L$ with respect to $G$.}
\end{figure}

At this point one can calculate $G$ and $L$; the results are shown in Fig.~\ref{fig2}. Notice the relative suppression of the $L$ form factor as compared to $G$; thus, based on Eq.~(\ref{alphaeq}), we expect that even in the region of intermediate momenta, where the difference reaches its maximum, the  two effective charges defined above will be comparable.

\begin{figure}[!t]
\includegraphics[scale=1.1]{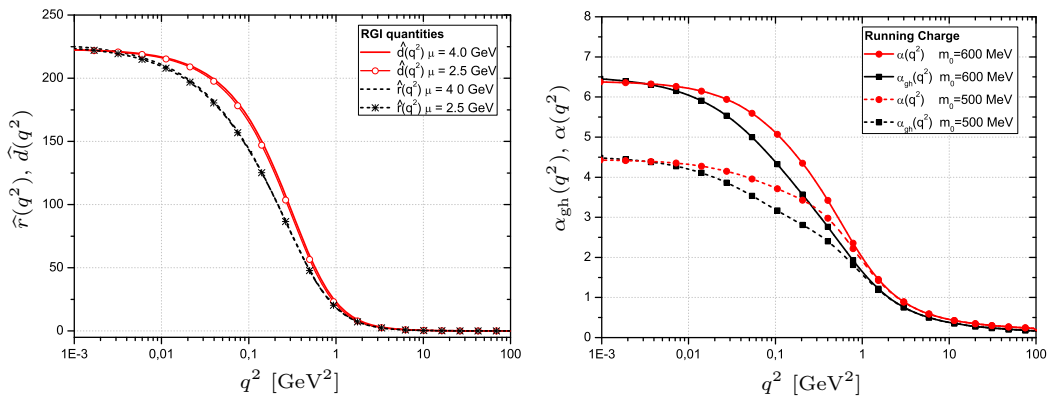}
\caption{\label{fig3}
{\it Left panel}: Comparison between the two RG-invariant products $\widehat{d}$ (solid line) and $\widehat{r}$ (dashed line); notice that there are two overlapping curves at different $\mu$ for each product. {\it Right panel}: 
Comparison between the QCD effective charge extracted from lattice data:  $\alpha$ (red line with circles) 
and $\alpha_{\mathrm{gh}}$ (black line with squares) for two different IR gluon masses: $m_0=500$ MeV (dashed) and $m_0=600$ MeV (solid).}
\end{figure}

Finally the RG invariant combinations~(\ref{RGIs}) can be constructed, and the corresponding effective charges calculated (Fig.~\ref{fig3}). Notice that has expected the differences between the two are in general small (around 10\%) making the two definitions practically indistinguishable.

\section{Conclusions and outlook}

\noindent In this talk we have shown that when applied to Yang-Mills theories, the BV formalism furnishes a set of very powerful identities that, when properly combined, allow to extract a great deal of information regarding the underlying non-perturbative dynamics of the theory. 

Future research directions include the application of the BV method to the Maximal Abelian Gauge and the study of the BFM in the presence of non-trivial backgrounds.\\

\noindent{\it Acknowledgments:} We thank the organizers of this Workshop for their warm hospitality and the stimulating atmosphere of the meeting.

\end{document}